\documentstyle [12pt] {article}
\title{A DE BROGLIE-LIKE WAVE IN THE PLANETARY SYSTEMS}

\author{Vladan Pankovi\'c$^{\ast,\sharp}$,Rade
Glavatovi\'c$^\diamond$, Marko Mandi\'c$^\sharp$, Vojislav Bo\v{z}i\'c - Sremac$^\sharp$\\
$^\ast$Department of Physics, Faculty of Sciences, 21000 Novi
Sad,\\ Trg Dositeja Obradovi\'ca 4. , Serbia, vdpan@neobee.net \\
$^\sharp$Gimnazija, 22320 Indjija, Trg Slobode 2a, Serbia \\
$^\diamond$ Military-Medical Academy, 11000 Belgrade, Crnotravska
17., Serbia\\}

\date {}
\begin{document}
\maketitle

\vspace {0.5cm} PACS number: 03.65.Ta, 96.35.-j \vspace {0.5cm}

\begin {abstract}
In this work we do an "interpolation" of Scardigli theory of a
quantum-like description of the planetary system that reproduces
remarkable Titius-Bode-Richardson rule. More precisely, instead of
simple, approximate, Bohr-like theory, or, accurate,
Schr$\ddot{o}$dinger-like theory, considered by Scardigli, we
suggest originally a semi-accurate, de Broglie-like description of
the planetary system. Especially, we shall propose a de
Broglie-like waves in the planetary systems. More precisely, in
distinction from Scardigly (which postulated absence of the
interference phenomena at planet orbits) we shall prove that,
roughly speaking, planets orbits equal a sum of natural numbers of
two types, large and small, of the de-Broglie-like waves. It is
similar to well-known situation in atomic physics by
interpretation of Bohr momentum quantization postulate by de
Broglie relation.
\end {abstract}
\vspace{0.5cm}
"Why you are miserable? - Because of my miseries\\
When Saturn packed my satchel I think\\
He put in these troubles - That's mad\\
You're his lord and you talk like his slave\\
Look what Solomon wrote in his book\\
"A wise man" he says, "has authority\\
Over the planets and their influence" -\\
I don't believe it, as they made me I'll be -\\
What are you saying? - Yes, that's I think -\\
I've nothing more to tell you - I'll survive without it-"\\

{\bf Francois Villon}, "The Debate Between One and Other Villon
Part,His Body And His Hart"

\newpage

In this work we shall do an "interpolation" of Scardigli theory of
a quantum-like description of the planetary system [1] that
reproduces remarkable Titius-Bode-Richardson rule [2]. More
precisely, instead of simple, approximate, Bohr-like (naïve
quantum-like) theory, or, accurate, Schr$\ddot{o}$dinger-like
(quantum-like) theory, considered by Scardigli, we shall suggest
originally a semi-accurate, de Broglie-like (quasi-classical-like)
description of the planetary system (without explicit Scardigli
consideration). Especially, we shall propose a de Broglie-like
waves in the planetary systems. More precisely, in distinction
from Scardigly (which postulated absence of the interference
phenomena at planet orbits) we shall prove that, roughly speaking,
planets orbits equal a sum of natural numbers of two types, large
and small, of de-Broglie-like waves. It is similar to well-known
situation in atomic physics by interpretation of Bohr momentum
quantization postulate by de Broglie relation (so that Bohr
electron orbit holds natural number of corresponding de Broglie
waves).

For a relatively small physical system with mass m, that stablely
rotates, with linear speed v, along a circumference with radius
$R$, about central, massive system $M$, for $M \gg m$, Scardigli
[1] suggest the following rules
\begin {equation}
      \frac {mv^{2}}{R} = G\frac {mM}{R^{2}}
\end {equation}
\begin {equation}
      \frac {J}{m} = vR =  S \exp[\alpha n] \hspace{1cm}   {\rm for}  \hspace{1cm}  n=1,
      2,3...
\end {equation}
where $\frac {J}{m} = vR$ represents the angular momentum of small
system over mass unit, while $S$ and $\alpha$ represent some
parameters independent of the natural number $n = 1, 2, 3, …$ .
(Originally, i.e. in [1], Scardigli uses denotation $\lambda$
instead of $\alpha$. But, since in our work we shall consider wave
length, usually denoted by $\lambda$, we change original Scardigli
denotation by $\alpha$. All other denotations in our work are
identical to Scardigli denotations.)

As it is not hard to see, given rules are, in some degree
conceptually similar to Bohr rules of electron motion within atom.
However, Scardigli pointed out that given similarity is not the
identity. He criticized some authors which use a non-modified Bohr
quantization form, i.e. that in (2) instead of $\exp[\alpha n]$
use simply $n$, since it yields $R$ as function of $n^{2}$ which
contradicts to empirical facts. Scardigli observed that
elimination of this contradiction, by conservation of non-modified
Bohr quantization form (simple term n), cannot be realized
physically plausiblely. Namely, it needs either non-consequitve
natural numbers or an ad hoc separation of Solar system in the
terrestrial and gigantic planets, any of which is physically
non-plausible.

Simple solution of Scardigli equations (1), (2) are
\begin {equation}
          v_{n} = \frac {GM}{S} \exp[-\alpha n] =  v_{1}\exp[-\alpha (n-1)] \hspace{1cm}   {\rm for}  \hspace{1cm}  n=1,
      2,3...
\end {equation}
and
\begin {equation}
          R_{n} = \frac {S^{2}}{GM} \exp[2\alpha n] = R_{1}\exp[2\alpha (n-1)] \hspace{1cm}   {\rm for}  \hspace{1cm}  n=1,
      2,3...
\end {equation}
where $R_{1} = \frac {S^{2}}{GM} \exp[2\alpha]$ and $v_{1} = \frac
{GM}{S} \exp[-\alpha]$ represent first planet, i.e. Mercury radius
and speed. Last term holds form identical to remarkable
Titius-Bode rule in Richardson form for planet distances [2],
under condition that $\frac {S^{2}}{GM}$ represents
Titius-Bode-Richardson parameter ${\it a}$. It implies
\begin {equation}
          S = ({\it a}GM)^{\frac {1}{2}}
\end {equation}
so that, as it is necessary, $S$ represents really a parameter
independent of m. (It can be added that Richardson form, (4), (5),
corresponds to usual form of Titius-Bode rule, only after
neglecting of the second order corrections.)

Further, Scardigli [1] suggest a more accurate, Schr$\ddot
{o}$dinger-like theory of the planetary systems, which, as its
simplification, holds mentioned, simple, Bohr-like theory.

Now, we shall consider originally a semi-accurate,
quasi-classical, de Broglie-like description of the planetary
systems (without explicit Scardigli consideration), that can be
considered as an "interpolation", between Bohr-like and Schr$\ddot
{o}$dinger-like theory. Especially, we shall propose a de
Broglie-like waves in the planetary systems (explicitly suggested
by Scardigli).

As it is well-known, de Broglie wave, corresponding to a quantum
particle with mass m and speed v, holds wave length $\lambda =
\frac {h}{mv}$, where $h$ represents Planck constant. It is
well-known too that Bohr momentum quantization rule, ad hoc
postulated within particle concept of quantum system,
$mv_{n}r_{n}=n\frac {h}{2\pi}$ for $n=1,2, …$ , can be simply
interpreted by given de Broglie relation. Namely, given relation
simply yields $2\pi r_{n}=n\frac {h}{mv_{n}}=n\lambda_{n}$   for
$n=1,2, …$ . It means that any electron orbit in Bohr atom holds
natural (positive integer) number of corresponding de Broglie wave
lengths. In other words, Bohr quantization rule can be considered
as a stability (equilibrium) condition for de Broglie waves
corresponding to electron. Simply speaking, without Bohr orbits,
electron de Broglie waves disappear, while, at Bohr orbits,
electron de Broglie waves stand conserved by means of the
interference effects (in a linear theory).

We shall suppose that in some degree similar situation exists
implicitly in mentioned Scardigli theory. Namely, we shall
suppose, according to (2), (3), (4),
\begin {equation}
     v_{n}R_{n} =  S \exp[\alpha n] = \frac {\sigma}{2\pi m} \exp[\alpha
     (n-1)] \hspace{1cm}   {\rm for}  \hspace{1cm}  n=1,
      2,3...
\end {equation}
where
\begin {equation}
      \sigma = S(2 \pi m) \exp[\alpha]           .
\end {equation}
It implies
\begin {equation}
     2\pi R_{n}= \frac {\sigma}{mv_{n}}\exp[\alpha (n-1)] \hspace{1cm}   {\rm for}  \hspace{1cm}  n=1,
      2,3...     .
\end {equation}
It, in conceptual similarity with de Broglie interpretation of
Bohr momentum quantization rule, suggests that, according to (7)
and (5), expression
\begin {equation}
     \lambda_{n} = \frac {\sigma}{mv_{n}}= \frac {2\pi S\exp[\alpha]}{v_{n}} =
     \frac {2\pi ({\it a}GM)^{\frac {1}{2}}\exp[\alpha]}{v_{n}} \hspace{1cm}   {\rm for}  \hspace{1cm}  n=1,
      2,3...
\end {equation}
can be considered as de Broglie-like wave length of a planet. It
is very interesting that given wave length, that depends of ${\it
a}$ and $M$, does not depend effectively of $m$, even if it is
defined by means of $m$.

However, in distinction from Bohr momentum quantization postulate
interpreted by de Broglie relation, expression
\begin {equation}
    2\pi R_{n}= \lambda_{n}\exp[\alpha (n-1)]  \hspace{1cm}   {\rm for}  \hspace{1cm}  n=1,
      2,3...
\end {equation}
, obtained by (8) and (9), cannot be, seemingly, simply
interpreted as a stability (equilibrium) condition (realized by
means of interference effects) within wave dynamics. Meanwhile,
(10) can be approximated by first order Taylor expansion of
$\exp[\alpha n]$ for n=1, 2, 3, … , which yields
\begin {equation}
     2\pi R_{n}= \lambda_{n} (1 + \alpha (n-1)) = \lambda_{n}  + (n-1)\lambda_{n red}  \hspace{1cm}   {\rm for}  \hspace{1cm}  n=1,
      2,3...
\end {equation}
where
\begin {equation}
     \lambda_{n red}= \lambda_{n}\alpha  \hspace{1cm}   {\rm for}  \hspace{1cm}  n=1,
      2,...        .
\end {equation}
The last term, i.e. $\lambda_{n red}$, can be considered as an
effective reduction, or, simply, small de Broglie-like wave length
corresponding to large de Broglie-like wave length, $\lambda_{n}$,
for $n=1, 2, 3, …$ . In other words, it can be supposed that here
is a non-linear wave dynamics whose stability (equilibrium)
condition (11) includes simultaneously two types of the waves. (It
would correspond to frequency mixing and similar phenomena in
non-linear optics too.) First one is usual, i.e. large de
Broglie-like wave, with de Broglie-like wave length $\lambda_{n}$,
for $n=1, 2, 3, …$ (9). Second one is corresponding, reduced, i.e.
small de Broglie-like wave with reduced, i.e. small de
Broglie-like wave length $ \lambda_{n red}$, for $n=1, 2, 3, …$
(12). Then condition (11) simply means that circumference of
$n$-th planetary orbit holds only one large de Broglie-like wave
length $\lambda_{n}$ and $n-1$ corresponding reduced, i.e. small
de Broglie-like wave lengths $\lambda_{n red}$ for $n=1, 2, 3, …$
. It is in a satisfactory conceptual similarity with Bohr momentum
quantization postulate interpreted by de Broglie relation.
Especially, as it is not hard to see, for the first planet (11)
has form (without small wave length) completely analogous to Bohr
quantization postulate for the first electron orbit.

Further, we shall consider two especial approximation limits of
(11).

Firstly, (11) turns out approximately in
\begin {equation}
     2\pi R_{n}\simeq \lambda_{n}    \hspace{0.5cm} {\rm for} \hspace{0.5cm} 1 \geq \alpha (n-1)  \hspace{0.5cm} {\rm or}  \hspace{0.5cm} n \leq \frac {1}{\alpha}+1  .
\end {equation}
It means that here circumference of $n$-th planetary orbit is
approximately equivalent to large de Broglie-like wave length
$\lambda_{n}$, for $n \leq \frac {1}{\alpha}$. Meanwhile, (13) has
form that is not completely analogous to form of de Broglie
interpretation of Bohr momentum quantization postulate (there is
no $n$ before $\lambda_{n}$ at right hand of (13)).

Secondly, (11) turns out approximately in
\begin {equation}
     2\pi R_{n}\simeq (n-1)\lambda_{n red}   \hspace{0.5cm}       {\rm for} 1  \hspace{0.5cm}< \alpha (n-1)   \hspace{0.5cm} {\rm or} \hspace{0.5cm} {\rm n} > \frac {1}{\alpha}+1 .
\end {equation}
It means that here circumference of $n$-th planetary orbit is
approximately equivalent to $n-1$ small de Broglie-like wave
length $\lambda_{n}$, for $n >\frac {1}{\alpha}+1$ . Obviously,
(14) has form almost completely analogous to form of de Broglie
interpretation of Bohr momentum quantization postulate. Namely
here $n$-th planet orbit holds approximately $n-1$ corresponding
reduced de Broglie-like wave lengths while $n$-th electron orbit
holds $n$ de Broglie wave length.

It can be observed that in Solar system $\alpha=0.2685$ that
implies $\frac {1}{\alpha} = 3.7239 +1 ˜ 5$. It practically means
that condition (13) is satisfied for terrestrial planets and
asteroids belt, while condition (14) is satisfied for gigantic
planets. Thus, we obtain separation of Solar system in terrestrial
(including asteroids belt) and gigantic planets in a reasonable
way, i.e. by two opposite limits of (11).

In conclusion we shall discuss obtained results. It can be
supposed [1], [2] (and references therein), that after formation
of Sun as massive central body rest of Sun system can be
satisfactorily treated as a real fluid-like system (consisting
really of the dust, gas etc.). Its non-linear dynamics can be, in
principle, accurately and completely described by some,
non-linear, e.g. modified, Schr$\ddot{o}$dinger-like equation or
similar partial differential equation. Solution of given equation
can accurately and completely predict how fluid density evolves
during time forming finally a discrete series of the splited parts
corresponding to permitted orbits of the future planets. (A planet
can be considered as the "condensate" of given fluid at permitted
orbits.) However, formulation as well as solution of given
equation is very complex and can be connected with many technical
as well as principal problems. For this reason an approximation of
given dynamics can be made by suggested de Broglie-like theory.
Instead of a complete dynamical evolution of the Sun system fluid
only final stable planets orbits can be proposed by corresponding
to Bohr-like, i.e. Scardigli, momentum quantization postulate. It
can be interpreted de Broglie-like waves (and interference rules)
as a stability condition in non-linear wave dynamics (with mixing
frequency, i.e. wave lengths effects). Roughly speaking future
planets orbits correspond to domains that hold the sum of one
large de Broglie-like wave length and integer number of small de
Broglie-like waves. In given domains interference effects yields
maximal amplification of the fluid density. In further
approximation orbits of the future terrestrial planets correspond
to domains that hold one large de Broglie-like wave length, on the
one side. On the other side, within the same approximation, orbits
of future gigantic planets correspond to domains that hold integer
number of small de Broglie-like wave lengths. Last case is almost
completely analogous to de Broglie interpretation of Bohr momentum
quantization postulate. Without mentioned domains interference
effects lead toward maximal decrease and final disappearance of
the fluid density.

\vspace{2.5cm}

{\Large \bf References}

\begin {itemize}

\item [[1]] F. Scardigli, {\it A Quantum-like Description of the Planetary Systems}, gr-qc/0507046, and references therein
\item [[2]] M. M. Nieto, {\it The Titius-Bode Law of Planetary Distances: its History and Theory} (Pergamon Press, Oxford, 1972)

\end {itemize}

\end {document}